\documentstyle[12pt]{article}
\begin{document}
\begin{flushright}
Univ. Freiburg\\
THEP 94/17\\
June, 1994\\
\end{flushright}
\vskip 1 cm
\begin{center}
{\large Some Features of Blown-Up Nonlinear $\sigma$-Models}\\
\vskip 2cm
H-B Gao$^*$ and H R\"omer\\
\vskip 1 cm
{\it
Fakult\"at f\"ur Physik\\
Universit\"at Freiburg\\
Hermann-Herder-Str. 3\\
79104 Freiburg, Germany\\}
\vskip 1 cm
{\bf \large Abstract}
\end{center}

In terms of the gauged nonlinear $\sigma$-models, we describe some results and
implications of solving the following problem: Given a smooth symplectic
manifold as target space with a quasi-free Hamiltonian group action, perform
the symplectic blowing up of the point singularity and identify the blow-up
modes in the corresponding (gauged) $\sigma$-model. Both classical and quantum
aspects of the construction are explained, along with illustrating examples
from the toric projective space and the K\"ahler manifold. We also discuss
related problems such as the origin of Mirror symmetry and the quantum
cohomologies.
\vskip 1.5 cm
\noindent $*$ {\small Permanent address: ZIMP, Zhejiang University, Hangzhou
310027, P.R. China}
\newpage

One of the characterizing properties of a string theory is the  duality
symmetry which exchanges the role of momentum and winding modes. It has
recently been discovered that duality symmetry can also be implemented
in the gauged version of some nonlinear $\sigma$-model \cite{rv}. This
appears to be somewhat mysterious. As the notion of the winding modes is
unique to strings, the result of \cite{rv} seems to suggest a new
geometric property undermined in the (gauged) nonlinear $\sigma$-models.

 In understanding the origin of the duality symmetry in nonlinear
$\sigma$-models, one may compare the case of the orbifold compactifications
of the string theory. Owing to the existence of the winding modes in the
string excitations, the orbifold singularities are automatically resolved
by adding the twisted sectors to the Hilbert space. The resulting low energy
effective theory is geometrically equivalent to a nonlinear $\sigma$-model
blown-up at the orbifold singular points. This fact is often phrased as
" string theory resovles (orbifold) singularity".  One  therefore  expects
that the  blowing up operation in a nonlinear $\sigma$-model would provide
additional structures responsible for the stringy properties such as duality
symmetry. Further evidence comes from recent studies of the topology changing
phenomenon in string theory \cite{tchang}.

Apart from the above "string-generated" motivation, there are more modest
purposes of studying blown-up nonlinear $\sigma$-models both from mathematical
and physical standpoints, among them we just mention two: The topological
$\sigma$-model \cite{w1} is originally proposed to capture some salient
features of Floer's theory of symplectic diffeomorphisms \cite{flo}, a basic
ingredient of the classification theory of the symplectic (as well as complex)
manifolds is the notion of birational equivalence. It is both necessary and
natural to include equivalence classes of the topological $\sigma$-models
obtained by blowing-up; In applying the Duistermaat-Heckman theroem \cite{dh}
to the singular case, one is led to consider the jump of the cohomology
class of the symplectic 2-form at the blowing up points \cite{au-gui}. The
natural invariant is the Duistermaat-Heckman polynomials appeared as the
coefficients of the expansion of the push-forward-by-moment-map symplectic
volume element in terms of the volumes of the subdivided convex polyhedral
sets. In a formulation of the blown-up $\sigma$-model, one is able to make
calculations which lead to quantities quite similar to the Duistermaat-Heckman
polynomials \cite{toapp}.

To describe what a blown-up nonlinear $\sigma$-model looks like, it is
necessary to recall the quotient constructions. One of them, which is called
symplectic reduction, will be our basic tool. So we begin by giving a brief
explanation.

Symplectic reduction is a procedure for obtaining quotient from a symplectic
manifold $M$ with Hamiltonian $G$-action, where $G$ is an arbitrary Lie group.
The action being Hamiltonian means simply that the interior product of the
vector field $X^a$ generating the action with the symplectic 2-form $\omega$,
is an exact 1-form:
$$i_{X^a} \omega=d\Phi^a, $$
where $\Phi^a$'s are scalar functions on $M$ the whole set of which is
called the moment map. When the
action is free of fixed points, one can go to the quotient of the level set
of the moment map, i.e. $\Phi^{-1}(m)$ at level $m$, by the isotropy subgroup
$G_m$ of the whole isometry group $G$, the resulting manifold $M'$ is a smooth
symplectic manifold with reduced symplectic form, and with dimension $2dimG_m$
smaller than $dimM$. There exists a construction of symplectic quotient in
terms
of nonlinear $\sigma$-model by gauging  the isometry corresponding to the
Hamiltonian $G$-actions. The procedure is roughly as follows. Let $g_{ij}$ be
the metric on $M$ which is compatible with the symplectic structure
$\omega=\omega_{ij}dx^i\wedge dx^j$, it means that $g_{ij}=\omega^k_i
\omega^l_j g_{kl}$. The action for the $\sigma$-model with target space $M$
is given by
$$S={1\over 2} \int d^2\sigma g_{ij}\partial_{\mu}x^i\partial_{\mu}x^j.$$
Let $\xi^i_a(x)$ be the Killing vectors generating the isometry
transformations,
then,
$$\nabla_{(i}\xi_{i)a}=0,$$
the $\sigma$-model fields transform according to
$$\delta x^i=\lambda^a\xi^i_a(x).$$
It is possible to make to transformation local by introducing a gauge field
$A^a_{\mu}$ which transforms as
$$\delta A^a_{\mu}=\partial_{\mu}\lambda^a(x) +f^a_{bc}A^b_{\mu}\lambda^c.$$

Consider the effective action of the form
$$S'(x)=-\ln Z(x),~~ Z(x)=\int DA D\eta \exp \{-{1\over 2}\int d^2\sigma
g_{ij}D_{\mu}x^iD_{\mu}x^j +\eta_a\Phi^a(x)\}$$
where $D_{\mu}x^i=\partial x^i-A^a_{\mu}\xi^i_a(x)$, $\Phi^a$ is a component
of the constraints specified by the moment map of the $G$-action, and
$\eta_a$ the Lagrangian multipliers. The effective action $S'$ therefore
describes a manifold $M'$ which is the symplectic quotient of $M$.

Although the expression for the $S'$ above is known to be affliated with
difficulties associated to functional integrations, there are some cases
where the tree level approximation seems to be exact as far as only
target space geometry is concerned, these include the $CP^n$ and $G(n,N)$
(complex Grassmannian) $\sigma$-models.

If the $G$-action on $M$ is not free, the naive symplectic reduction fails
because the resulting quotient ceases to be a manifold, but rather a singular
space. The mathematical studies of the singular symplectic reduction have
revealed some interesting structures such as stratification \cite{lerm},
but global properties of the general singular symplectic reduction remain
largely unclear. Quite recently, singular reduction has been employed in
a beautiful work by Guillemin and Sternberg \cite{gs89} in their formulating
the notion of the symplectic cobordism. In a sense, the results of \cite{gs89}
suggest a role of singular symplectic reduction as a subsititute for
performing blowing up of singularities in a Hamiltonian $G$-space. In our
previous paper \cite{me} we have explained how the symplectic blowing up of
Guillemin-Sternberg can indeed been realized in terms of the gauged nonlinear
$\sigma$-models. The basic construction involves viewing the operation  of
symplectic blowing up as a combination of the local and global reductions.
Let us review the main points in \cite{me}. Recall the geometric content of
symplectic blowing up \cite{grom}. Any point $x_0$ in a $2m+2$ dimensional
symplectic manifold $(M,\omega)$ has a small neighborhood ${\cal U}_0\subset
M$ which is isometric to $C^{m+1}$ with standard linear symplectic form
$\omega_0$. We cut open $M$ along  a solid ball $B$ of radius $\epsilon$
in ${\cal U}_0$ and
collapse the boundary $\partial B \sim S^{2m+1}$ of $B$ into a complex
projective space $(CP^m, \omega_1)$. On $CP^m$ there is a canonical line
bundle $p: L\rightarrow CP^m$ which is constructed from simple incidence
relation as subset of the product $C^{m+1} \times CP^m$, i.e. $L=\{(x,y)
\in C^{m+1}\times CP^m \vert xy-yx=0\}$. the projection of $L$ onto the
first factor $\sigma: L\rightarrow C^{m+1}$ is a holomorphic map which has the
following properties. $\sigma$ restricted to the complement of the zero
section of $L$ is bi-holomorphic onto $C^{m+1}-\{0\}$, and
$\sigma^{-1}(0)=CP^m$.
By attaching $\sigma^{-1}(0)=CP^m$ to $M-B$, one obtains the blown-up
symplectic manifold $M'$ which is smoothly diffeomorphic to the connected
sum $M\#CP^{m+1}$ endowed with a well-defined symplectic form
$$\omega'=\sigma^*(\omega_0)+\epsilon p^*(\omega_1).$$

In our $\sigma$-model relazation of symplectic blowing up, the process of
cutting open $M$ is replaced by performing the normal coordinate expansion
at the fixed point $x\in M$ of the quasi-free\footnote{A $G$-action is said
quasi-free, if the stabilizer groups are either empty or the whole $G$.
This has been imposed mainly to simplify presentation, more general cases
can in principle be handled similarly. However, the case of singular
submanifold deserves another study which involves blowing up along
submanifold and knowing its normal forms. The gauged $\sigma$-model
realization of the later case requires auxilliary fields coming from
the coadjoint orbit of $G$.} $G$-action, and
that of collapsing boundary of $B\subset {\cal U}_0$ is replaced by gauging
of an Abelain $U(1)$ isometry of a linear $\sigma$-model resulted from
normal coordinate expansion. This is the local reduction advocated to above.
The global reduction turns out to rely on a use of symplectic diffeomorphism
whose fixed point does not coincide with the fixed point of the Hamiltonian
$G$-action. The role of this symplectic diffeomorphism is to glue the blow-up
exceptional divisor back to the complement of the blown-up point in the
nonlinear $\sigma$-model. Thus we arrive at the conclusion that the blown-up
$\sigma$-model is a sum of two gauged nonlinear $\sigma$-models related by
a symplectic diffeomorphism. The classical target space geometry so obtained
is a connected sum $M\#CP^m$, with a well-defined symplectic form.
One may view the process as depicted in Figure 1.

Let us make the following remarks regarding the above result.
\begin{enumerate}
\item The parameter $\epsilon$ is to be interpreted as volume of the
exceptional divisors by Poincar\'e duality. When $\vert \epsilon\vert$
is small, both $+\epsilon$ and $-\epsilon$ correspond to regular levels
of the $U(1)$ moment map upon which the symplectic reductions are performed.
As $\epsilon$ approaches zero form either sides, either volume of the
exceptional divisor becomes infinitely small and finally the exceptional
divisor disappears, or reversely the zero limit of the parameter $\epsilon$
signals a nontrivial exceptional divisor, representing the blowing-up
mode in the corresponding $\sigma$-model.
\item It is interesting to explore the effects of the instanton associated
with the exceptional divisor of the blowing up. One may interpret this
blowing up instanton as a small perturbation  with parameter $\epsilon$
of the instantons which exist before blowing up. The point is that, since
$\epsilon$ can take positive and negative values, a smooth change of
topology may occur in which instantons emerge or disappear at some ponts
in the manifold. This has drastical implications for the quantum properties
of the model, such as spontaneous symmetry breaking and the solvability of
the model in large $N$. This kind of study is being carried out
by the author \cite{me2}.
\end{enumerate}

In what follows we look at the properties of the blowning up operation
in some special nonlinear $\sigma$-models.

\noindent {\it Toric $\sigma$-model}

A toric manifold associated with an integral or rational polyhedron can be
obtained as a symplectic reduction of $C^N$ by the Hamiltonian action of the
subtorus of $T^N_C=(C^*)^N$, at a regular level of the corresponding monemt
map (see \cite{au} for an explanation). Obviously a toric $\sigma$-model
(i.e. a $\sigma$-model with target space being a toric manifold) can be viewed
as a suitable symplectic reduction of the linear $\sigma$-model on $C^N$.
It is a nontrivial fact that some
\begin{center}
{\small {\bf Fig. 1.} A connected sum resulted from blowing up.} \end{center}
\newpage
\noindent  Hamiltonian subtorus actions on the toric
manifold can have fixed points when the image of these points are exactly the
vertices of the convex polyhedron.  Let us take the simplest example of
 $CP^2$, constructed as a toric manifold whose associated polyhedron is the
 standard 2-simplex, i.e. a triange $\Delta \subset R^2$. If $e_1, e_2$
 denote the basis vectors of $Z^2 \subset R^2$, which are two edge vectors
 of $\Delta$, we can form a fan $\Sigma$ whose 1-skeletons (edges of the
 2-cones) are
 all of the form $tx_i, ~~0\leq t <\infty, ~ i=1,2,3$, where $x_i=e_i, x_3
 =-(e_1+e_2)$ (see Figure 2). One can take $x_i$ to be the basis vectors in
 $Z^3$, thus there exists a natural map $Z^3 \rightarrow Z^2,~~{\bf x}\mapsto
 {\bf z}$ which induces the corresponding map $R^3 \rightarrow R^2$ and the
 quotient map $T^3 \rightarrow T^2 \rightarrow 0$ with kernel $S^1$. The
 realization of $CP^2$ as symplectic reduction of $C^3$ is carried out by
reducing $C^3$ by the (smooth) Hamiltonian action of this $S^1$. From this
construction it is obvious that  $T^2 \subset T^3$ acts on $CP^2$ in a
Hamiltonian fashion and the image of $CP^2$ under its moment map is exactly
$\Delta$.

It is well-known that the nonlinear $\sigma$-model of $CP^2$ can be described
by a gauged $\sigma$-model with the gauge field
$$A_{\mu}={i\over 2}\sum^3_{i=1}{\bar z}^i\partial_{\mu}z^i
-z^i\partial_{\mu}{\bar z}^i,$$
and the action of the form $\int D_{\mu}z^i \overline{D_{\mu}z^i},~ D_{\mu}=
\partial_{\mu}-iA_{\mu}$. The integral of the symplectic form over a homology
cycle in $CP^2$ equals a topological invariant $1/2\pi\int \epsilon_{\mu\nu}
\partial_{\mu}A_{\nu}$ which is the first Chern number of the tangent bundle.
The $CP^2 ~\sigma$-model has a global $SU(3)$ invariance, of which the maximal
torus $T^2$ acts in the Hamiltonian fashion. We know that this $T^2$ action
is not free at some points whose image under the moment map are the vertices
of $\Delta$. This may cause serious problems in the quantum theory, eventhough
it is harmless classically, as far as one does not perform the quotient
(which would be a
\begin{center} {\small {\bf Fig. 2.} A convex polyhedron and its dual fan.}
\end{center}
\newpage
\noindent point in this case). A possible resolution is to blow up
the fixed point on $CP^2$.
Given the combinatoric data determining $\Sigma$ or $\Delta$, it is easy to do
blowing up in toric $\sigma$-model. For this purpose let us observe that, the
$CP^2$ $\sigma$-model is equivalent to the $SU(3)$ invariant chiral model
defined by the action desity
$$S={1\over 2} tr \partial_{\mu}\phi\partial_{\mu}\phi$$
where $\phi$ is a $3\times 3$ Hermitian matrix of trace zero. Indeed, let the
torus $T^3$ act on the space of $\phi$ matrices by assyning to the diagonal
elements of $\phi$ by the following set of eigenvalues $(\lambda_1,
\lambda_2, \lambda_3)=({1\over 3},{1\over 3}, -{2\over 3})$, this determines
$\phi$ uniquely from the eigenvectors $z=(z^1, z^2, z^3)$ as follows
$$\phi_{ab}=\delta_{ab}/3- z_a{\bar z}_b,~~~\vert z\vert^2=1.$$
It can be proved that
$$S={1\over 2} tr \partial_{\mu}\phi\partial_{\mu}\phi=\sum_a(\partial_{\mu}z^a
-iA_{\mu}z^a)(\partial_{\mu}{\bar z}^a+iA_{\mu}{\bar z}^a)$$
where $A_{\mu}$ is as before. From the second last equation, it is clear that
the diagonal elememts of $\phi$ lie in the convex polyhedron $\Delta$ which is
the image of the $T^2\subset T^3$ moment map. Now recall that blowing up toric
manifold at a fixed point just correspond to truncation of a vertex in $\Delta$
(see Figure 3). The dual fan is shown in Figure 4.

\vskip 4.0cm
\begin{center} {\small {\bf Fig. 3}. The blow up as truncation of
{}~~~~ {\bf Fig. 4}. A fan representing }\\
{\small ~~~ vertex of polyhedron. ~~~~~~~~~~~~~~~~~~~ blown-up of $CP^2$.}\\
\end{center}

Using the basic toric
methods, one concludes that the $CP^2$ model blown up at a point is obtained
by enlarging the rank of $\phi$ by one in such a way that it becomes a block
form with the $T^2=S^1\times S^1$ action on the last entry of diag($\phi$)
twisted by a sufficiently large integer $a$ (with $\epsilon\sim 1/a$).
The symplectic reductions which
lead to the blown-up $CP^2$ is reminiscent of our general senario, but is in
this case much easier. The benefit of working with the toric convex data
is that it becomes apparent how to identify the fixed points and the blow-up
exceptional divisors (see Figure 5). The later are closely related to the
important object, rational curves on $CP^2$.  We refer to \cite{me2} for
detailed account of the toric $\sigma$-models along with more results on
instanton analysis associated to the holomorphic curves in toric manifolds.
Our next example deals with holomorphic K\"ahler quotient rather than its
equivalent symplectic  analogue.
\vskip 3cm
{\small {\bf Fig. 5}. Fixed points and holomorphic curves in the fan of
blown-up $CP^2$.}

\noindent {\it N=2 Supersymmetric $\sigma$-model}

In this case, there exists a general procedure \cite{3,rv}
 of performing the $N=2$ quotient
by gauging the (holomorphic) isometries of the $N=2$ superspace action of the
form
$$S=\int D_+D_-{\bar D}_+{\bar D}_-K(\Phi, {\bar\Phi}, \Lambda, {\bar\Lambda}
), $$
with arbitrary chiral and twisted chiral superfield multiplets $\Phi, \Lambda$.
The gauged action takes the general form of a new K\"ahler potential $K'$ which
is the original potential $K$ with $\Phi$ and $\Lambda$ minimally coupled to
some gauge vector multiplet $V$, plus terms which are trivially gauge
invariant,
such as the Fayet-Iliopoulos terms which must be included when the isometry
group contains $U(1)$ subgroups. In the same spirit as the bosonic
$\sigma$-model and its symplectic reduction, we can carry out the symplectic
blowing up using the general method of \cite{me}. The basic ingredients are a
superspace analogue of the normal coordinate expansion on the one hand,
and the identification (and the interpretation) of the blowing up parameter
$\epsilon$  as the coupling constant in front of the Fayet-Iliopoulos terms,
on the other hand. We will not describe these points here. But the picture of
the blown-up $N=2$ $\sigma$-model is clear: to each $N=2$ supersymmetric
$\sigma$
model, arising from gauging an appropriate holomorphic isometry group, at any
(isolated) configuration which is the fixed point of a subgroup of the
isometry group, one can associate a gauged version of the $N=2$ linear
$\sigma$-model. The resulting $N=2$ model has a bosonic sector which looks
exactly like the symplectic blowing up of the type described before.

We would like to make some observations concerning  related issues of quantum
cohomology and mirror symmetry which have attracted much attention recently.

It is believed that some $N=2$ nonlinear $\sigma$-models admit equivalent
Landau-Ginsburg descriptions. For the $N=2$ Landau-Ginsburg model of the
complex Grassmannian $G(m, m+n)$ \cite{int},  the
K\"ahler form $X^{(1,1)}$ serves as perturbation of the ordinary cohomology
ring. With the potential perturbed to the form
$$W^m_{m+n+1}(X^{(i,i)})+(-1)^m \beta X^{(1,1)},$$
the corresponding quantum cohomology ring is
$$X^mY^n=\beta,$$
where $\beta$ can be taken as the (exponential of the) volume of the
exceptional divisor corresponding
 to an instanton. If the instanton comes from our blow up operation, its
 volume can be both positive and negative. We thus conclude that,
 with blow up modes included, some (derived) subring of the quantum cohomology
 ring becomes $Z^2$-graded.
It is an interesting question to find the physical degree which is responsible
to this grading.

Finally, the origin of the mirror symmetry seems to be connected \cite{vafa}
to the Abelian duality symmetry mentioned at the beginning. If the blown-up
$\sigma$-model indeed explains the origin of the duality symmetry, one would
be a step closer to understanding the origin of the mirror symmetry. Of course
much work must be done before any definite assertion can be made. In this
reagard, however, the following remark might be helpful. As Roan has shown
in \cite{roan} (see also recent work of \cite{bat}), some mirror pairs found
in \cite{12} admit interpretation as living over the dual lattices determining
the corresponding manifolds. Our treatment of the toric example above suggests
that a chain of chiral fields defined on the 2-dimensional lattice whose links
coincide with some subset of the fan $\Sigma^{(1)}\subset \Sigma$, serves
to provide an analogue of the mirror map for a pair of toric manifolds.
Details will be reported elsewhere.

\vskip 0.5cm
\noindent {\bf \large Acknowledgement}

One of us (HBG) thanks
the Alexander von Humboldt Stiftung for its kind support.

\end{document}